\documentclass[conference]{IEEEtran}

\usepackage{geometry}
\usepackage{todonotes}
\usepackage[utf8]{inputenc}
\usepackage{amsmath,amssymb,amsfonts}
\usepackage{textgreek}
\usepackage{graphicx}
\usepackage{color,bm}
\usepackage{wasysym}
\usepackage{tikz}
\usepackage{cite}
\usepackage{array,multirow}
\usepackage{float}
\usepackage{subfig}
\usepackage{balance}
\usepackage{hyperref}
\hypersetup{hidelinks}
\usepackage{mathtools, nccmath}
\DeclareGraphicsExtensions{.tif}
\usepackage[linesnumbered,ruled,vlined]{algorithm2e}
\usepackage{amsmath} 
\usepackage{algorithmic,algorithm2e}

\ifCLASSINFOpdf
\else
\fi
\begin{document}

\title{\LARGE \textbf{Augmented Reality-Based Advanced Driver-Assistance System for Connected Vehicles}}
      
\author{
Ziran Wang\IEEEauthorrefmark{1},
Kyungtae Han,
and Prashant Tiwari\\

Toyota Motor North America R\&D, InfoTech Labs, Mountain View, CA, USA\\
\{{\href{mailto:ziran.wang@toyota.com}{ziran.wang}}, {\href{mailto:kyungtae.han@toyota.com}{kyungtae.han}}, {\href{mailto:prashant.tiwari@toyota.com}{prashant.tiwari}}\}@toyota.com
}

\markboth{IEEE Transactions on Intelligent Vehicles,~Vol.~xx, No.~x, JUL~2020}%
{Shell \MakeLowercase{\textit{et al.}}: Bare Demo of IEEEtran.cls for IEEE Journals}
\maketitle
\begin{abstract}
With the development of advanced communication technology, connected vehicles become increasingly popular in our transportation systems, which can conduct cooperative maneuvers with each other as well as road entities through vehicle-to-everything communication. A lot of research interests have been drawn to other building blocks of a connected vehicle system, such as communication, planning, and control. However, less research studies were focused on the human-machine cooperation and interface, namely how to visualize the guidance information to the driver as an advanced driver-assistance system (ADAS). In this study, we propose an augmented reality (AR)-based ADAS, which visualizes the guidance information calculated cooperatively by multiple connected vehicles. An unsignalized intersection scenario is adopted as the use case of this system, where the driver can drive the connected vehicle crossing the intersection under the AR guidance, without any full stop at the intersection. A simulation environment is built in Unity game engine based on the road network of San Francisco, and human-in-the-loop (HITL) simulation is conducted to validate the effectiveness of our proposed system regarding travel time and energy consumption.
\end{abstract}

\IEEEpeerreviewmaketitle

\section{Introduction} \label{sec:Intro}
The emergence of connected vehicle technology during the past decades brings many new possibilities to our existing transportation systems. Specifically, the level of connectivity within our vehicles has greatly increased, allowing these ``equipped'' vehicles to behave in a cooperative manner not only among themselves through vehicle-to-vehicle (V2V) communication, but also with other transportation entities through vehicle-to-infrastructure (V2I) communication, vehicle-to-cloud (V2C) communication, vehicle-to-pedestrian (V2P) communication,  etc., namely vehicle-to-everything (V2X) communication.

Many research works have widely studied various aspects of the connected vehicle systems, such as communication, perception, localization, planning, and control, where each of them handles one or several tasks in the system \cite{wang2020asurvey}. The latter four aspects are often studied in the autonomous driving domain as well, and the concept of connected and automated vehicles (CAV) emerges where vehicles can conduct cooperative automation maneuvers together. However, it is expected that full automation of our transportation systems will not happen anytime soon, due to the hurdles in both the technical side and the liability side. Therefore, during the transition from no automation to full automation in the mixed traffic environment, human-driven connected vehicles will play a crucial role given the rich information they can share through V2X communication, as well as the ability to cooperate with other human-driven connected vehicles or CAVs.

Therefore, the importance of studying the topic of human-machine cooperation arises, as the driver of a connected vehicle needs to know how to correctly interact with the vehicle to maximize its full advantages \cite{monreal2016human}. One critical aspect of this topic is to design the human-machine interface (HMI) of the advanced driver-assistance system (ADAS), so that the information received through V2X communication can be visualized to the driver, and guides him/her to driver the vehicle in a safer, more efficient, and more comfortable way.

Connected vehicles have been well researched regarding the planning and control aspects, even with numerous field implementations conducted by real mass-produced vehicles. However, most of them designed the HMI of their ADAS as a simple visualization tool of the connected vehicles' planning and control modules, such as the driver guidance on the connected eco-driving system \cite{wang2019early}, or the driver vehicle interface design by 
U.S. Department of Transportation \cite{campbell2016human}. Only a relatively small portion of them addressed the issue from the HMI perspective, which designed the connected vehicle's planning and control modules according to the pattern of the HMI, making the integrated ADAS more informative while also intuitive for the driver to operate.

In this study, we propose an ADAS for connected vehicles using augmented reality (AR) as the HMI, which overlays the guidance information on driver's field-of-view through the windshield. A specific use case of unsignalized intersection is studied, where connected vehicles (including CAVs) can cooperate with each other to cross intersections without any full stop, largely increasing the time efficiency and energy efficiency of vehicles. A slot reservation planning algorithm and a feedforward/feedback control algorithm are developed to serve the AR HMI of the ADAS. Unity game engine is used to model the proposed system, and human-in-the-loop (HITL) simulation is conducted to validate the effectiveness of this system in the unsignaliezd intersection use case.  

The remainder of this paper is organized as follows: Section \ref{sec:Problem} introduces the problem statement of this study in a greater details. Section \ref{sec:Method} develops different modules of this ADAS on connected vehicles, including the AR HMI design for drivers, the slot reservation planning algorithm, and the feedforward/feedback control algorithm. Section \ref{sec:Sim} conducts the modeling and evaluation works of this ADAS in Unity game engine, with results in HITL simulation showing the effectiveness of the system. Finally, the paper is concluded with some future directions in section \ref{sec:Con}.

\begin{figure*}[ht!]
    \centering
    \includegraphics[width=2.0\columnwidth]{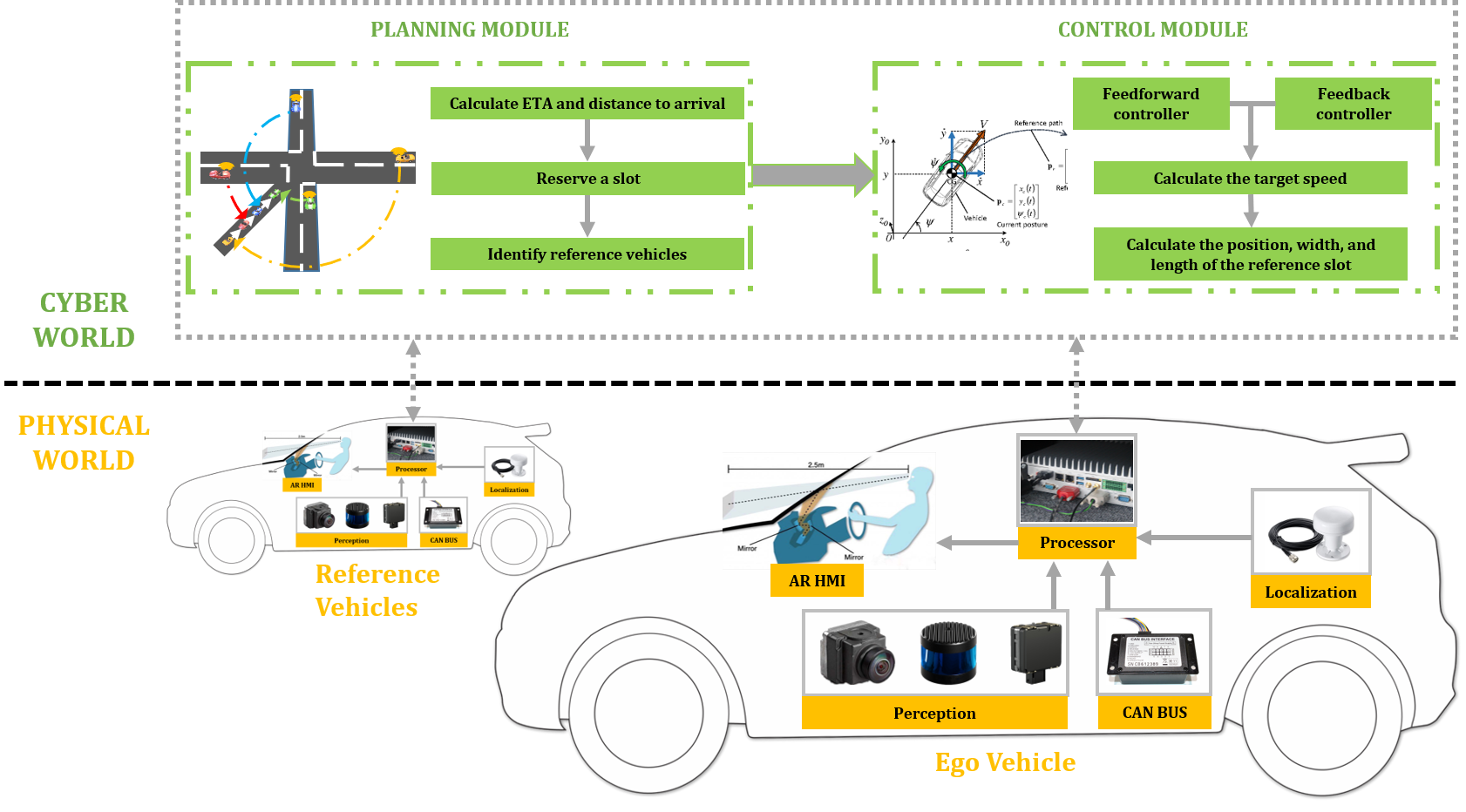}
    \caption{System architecture of the proposed AR-based ADAS for connected vehicles}
    \label{Arc}
\end{figure*}

\section{Problem Statement} \label{sec:Problem}
In this study, an ADAS is designed for connected vehicles, which includes various modules such as communication, localization, perception, planning, control, and AR HMI. Although every module is essential to the overall system architecture, we focus on the latter three in this study. Connected vehicles in this study can either be driven by human drivers with AR HMI, or driven by automated controllers as CAVs. A Digital Twin (i.e., cyber-physical) architecture is adopted for connected vehicles in this study, where all connected vehicles in the physical world are connected through the cyber world. The proposed cooperative maneuvers among connected vehicles does not specify or require any specific communication technology, which means vehicles can potentially be connected with the cyber world through Dedicated Short-Range Communications \cite{kenney2011dedicated}, Cellular Vehicle-to-Everything (C-V2X) \cite{chen2017vehicle}, or a combination of both.

The general architecture of the proposed ADAS can be illustrated as Fig. \ref{Arc}. In the physical world, connected vehicles are equipped with different hardware modules that can provide information of themselves and the surrounding traffic. For example, CAN BUS provides speed, acceleration, and many other detailed information of the ego vehicle, while the localization module provides its coordinate information. The perception module, on the other hand, provides surrounding information to the ego vehicle, such as road geometry, detected objects, and traffic condition. All these information is fed into the processor of the ego vehicle, which processes the data locally and sends to the cyber world through V2X communication. The processor also receives information from the cyber world and propagates to the AR HMI for the driver-assistance purpose. It should be noted that, this study does not focus on the hardware setup of the ADAS, as long as the necessary information can be gathered and provided to the planning, control, and HMI modules.

In the cyber world, the planning and the control modules play important roles in the overall system. The planning module schedules different connected vehicles before they conduct cooperative maneuvers (e.g., crossing the unsignalized intersections in this use case), allowing connected vehicles to identify their desired motions. The control module calculates the particular control commands to allow vehicles to achieve their desired motions, and executed either by human drivers or automated controllers.

Compared to previous studies on AR or ADAS of connected vehicles, the major contributions of this study are listed below:

\begin{itemize}
    \item \textit{Design the planning and control modules of the ADAS to better serve the HMI, making the system human-centered.} The HMIs in most existing ADAS simply visualize the information derived from other modules of their systems, such as visualizing the traffic light information received through V2X communication \cite{wang2019early, altan2017glidepath, conceicao2013virtual}, providing collision warning messages  \cite{monreal2016human, monreal2010theseethrough}, or displaying downstream traffic-related messages \cite{bertini2006dynamics, nygardhs2007variable}. In this study, however, we design the HMI using AR in a slot reservation format, and then develop the planning and control modules according to that. The HMI is not a natural output of this ADAS, but the basis of this ADAS that facilitates the human-centered human-machine cooperation.
    
    \item \textit{Adopt AR to overlay the guidance information on top of the traffic environment from the driver's field-of-view, providing more intuitive guidances.} Instead of adopting the AR concept to show traffic information on a separate display \cite{quinlan2010bringing, feng2018anaugmented}, or to show some simple information on the head-up display \cite{wang2020driver}, we adopt AR to overlay the guidance information on top of the traffic environment. This HMI provides a more intuitively meaningful indication of reference connected vehicles' presence and current status, and better assists the driving maneuver of the driver.
    
    \item \textit{Enable the cooperative maneuvers among connected vehicles through AR HMI, and validate the system through game engine modeling and HITL simulation.} Most of the existing automotive AR HMI designs are only for ego vehicle's maneuvers, such as navigation, speed visualization, and driving mode visualization \cite{continentalAR, vwAR}. In this study, we leverage V2X communication to allow connected vehicles to conduct cooperative maneuvers with the assistance of AR HMI. Not only do we design the AR HMI together with its associated planning and control modules in the system, but also conduct modeling and simulation with Unity game engine, which validates its effectiveness in an unsignalized intersection use case. 
\end{itemize}

\section{Augmented Reality for Advanced Driver-Assistance System} \label{sec:Method}
\subsection{Use Case of Unsignalized Intersections}
By design, an intersection is a planned location where vehicles traveling from different directions may come into conflict, and its functional area extends upstream and downstream from the physical area of the crossing streets. Traffic signals have been playing a crucial role in achieving safer performance at intersections, which can reduce the severity of crashes if operated properly \cite{fhwa2014signalized}. However, the addition of unnecessary or inappropriately-designed signals have adverse effects on traffic safety and mobility. In addition, the dual objectives of safety and mobility introduce trade-offs in many cases.

Therefore, the designs of unsignalized intersections emerge during recent years, which take advantage of the connected vehicle technology. Specifically, approaching vehicles can be assigned specific sequences by the proposed planning/scheduling algorithms through V2X communication, and their motions will be controlled by automated controllers or drivers with guidance information. Most existing works in this use case assume automation of connected vehicles \cite{dresner2008amultiagent, neuendorf2004thevehicle, jin2013platoon, xu2018distributed}, namely all vehicles in the system are CAVs. However, in this study, we aim to develop an AR-based ADAS that allows human-driven connected vehicles to perform the cooperative maneuvers at unsignalized intersections. This enables a more realistic application in this use case, because not all vehicles will become automated vehicles in the very near future, and there will definitely be a transition period of mixed traffic environment (that has both human-driven and automated vehicles).

In this case study, since we focus on the effectiveness of our proposed AR-based ADAS, some reasonable specifications are made regarding the settings of the use case: 1) The ego connected vehicle can receive information regarding vehicles coming from other directions of this intersection, either directly through V2V communication, or indirectly through V2I communication or V2C communication; 2) Except for the ego vehicle, not all other vehicles are required to be connected vehicles. In the case that certain vehicles do not have connectivity, the perception sensors equipped on their surrounding connected vehicles or on the intersection infrastructures can measure the information of those unconnected vehicles, and share it with the ego vehicle; 3) No vulnerable road users (e.g., pedestrians, bicycles, etc.) are considered in this use case.

\subsection{Design of the Augmented Reality Human-Machine Interface}
In this study, we propose an AR HMI that guides the driver to drive the connected vehicle and cross the unsignalized intersections with other connected vehicles. The information that needs to be visualized to the driver through the AR HMI is regarding vehicles coming from other directions of the intersections. We propose a slot reservation methodology to strategically allocate different vehicles with slots upon their approaches to the intersection. While the details regarding the slot reservation will be covered in the next subsection under the planning module, the design of the reserved slots on the AR HMI is introduced here. 

\begin{figure}[ht!]
    \centering
    \includegraphics[width=1.0\columnwidth]{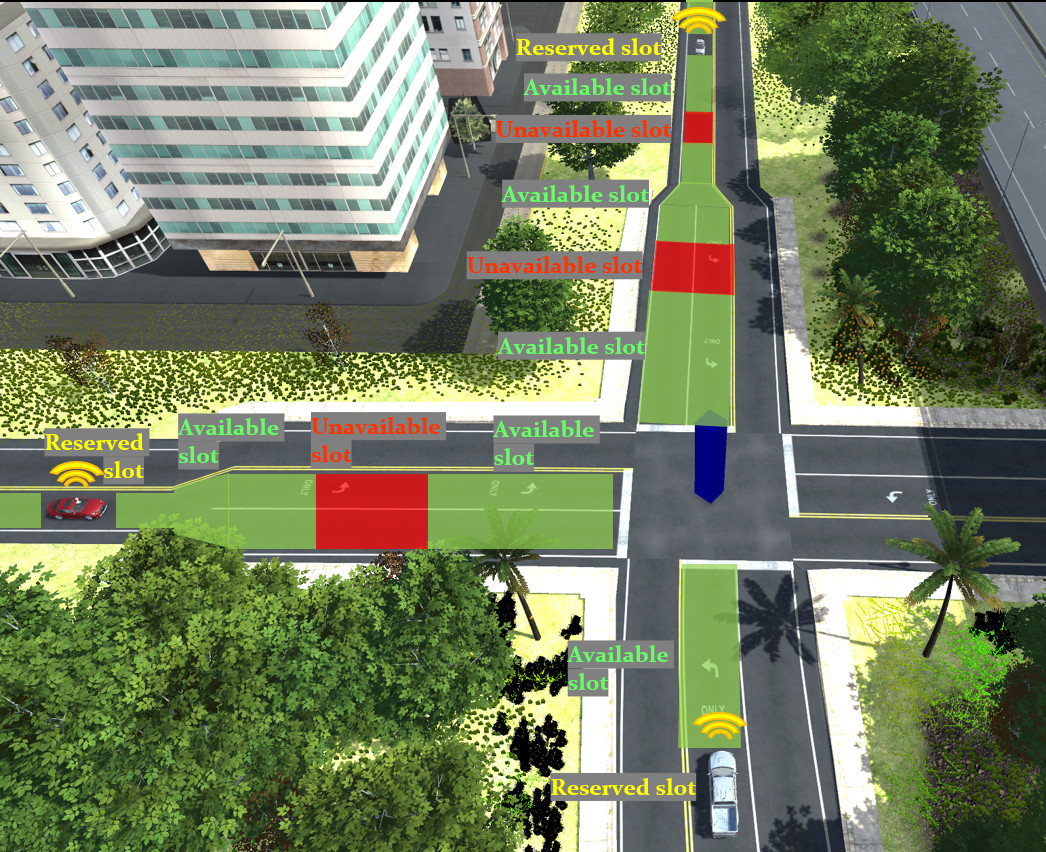}
    \caption{Illustration of the slot reservation concept for connected vehicles crossing an unsignalized intersection}
    \label{AR}
\end{figure}

A simple example of the unsignalized intersection is illustrated in Fig. \ref{AR}, where three connected vehicles are approaching the intersection from three directions. Once an ego vehicle gets assigned a slot, its information will be shared with its conflicting vehicles, whose paths have conflicting points with the ego vehicle's path. The slot reserved by the ego vehicle will then be shown to the drivers of conflicting vehicles as a red ``unavailable slot'' through AR HMI, so those drivers can control their vehicles to stay in the green ``available slots''. It needs to be noted that all slots are not stagnant, which are dynamically updated according to the status changes of their associated vehicles.

\begin{figure}[ht!]
    \centering
    \includegraphics[width=1.0\columnwidth]{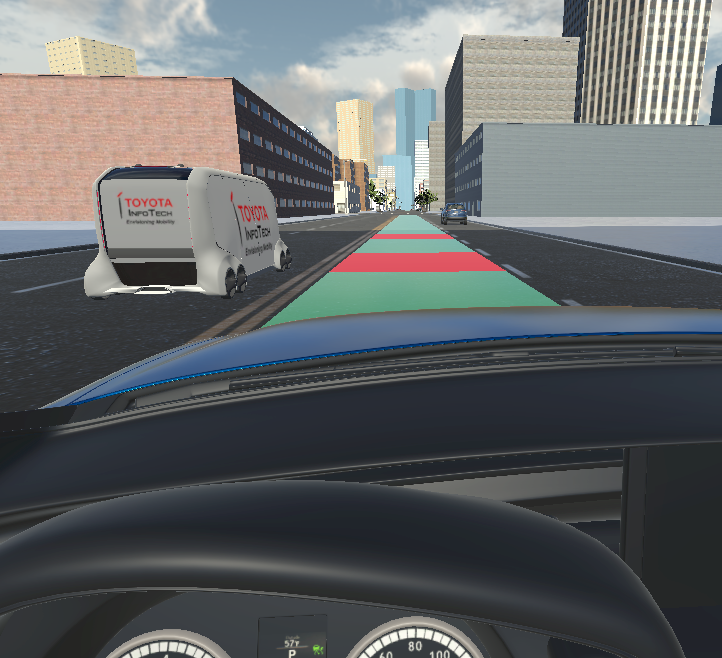}
    \caption{Driver's field-of-view of the AR HMI}
    \label{AR2}
\end{figure}

The high-level concept of the AR HMI is illustrated in Fig. \ref{AR}, while an example from the driver's field-of-view is shown in Fig. \ref{AR2}. The ego vehicle is approaching an unsignalized intersection, where two unavailable red slots are visualized on the HMI, denoting there are two conflicting vehicles coming from other directions of the intersection. The driver of this ego vehicle needs to control the vehicle to keep in the green available slots, so it can avoid collision while crossing the intersection.

\begin{figure}[ht!]
    \centering
    \includegraphics[width=1.0\columnwidth]{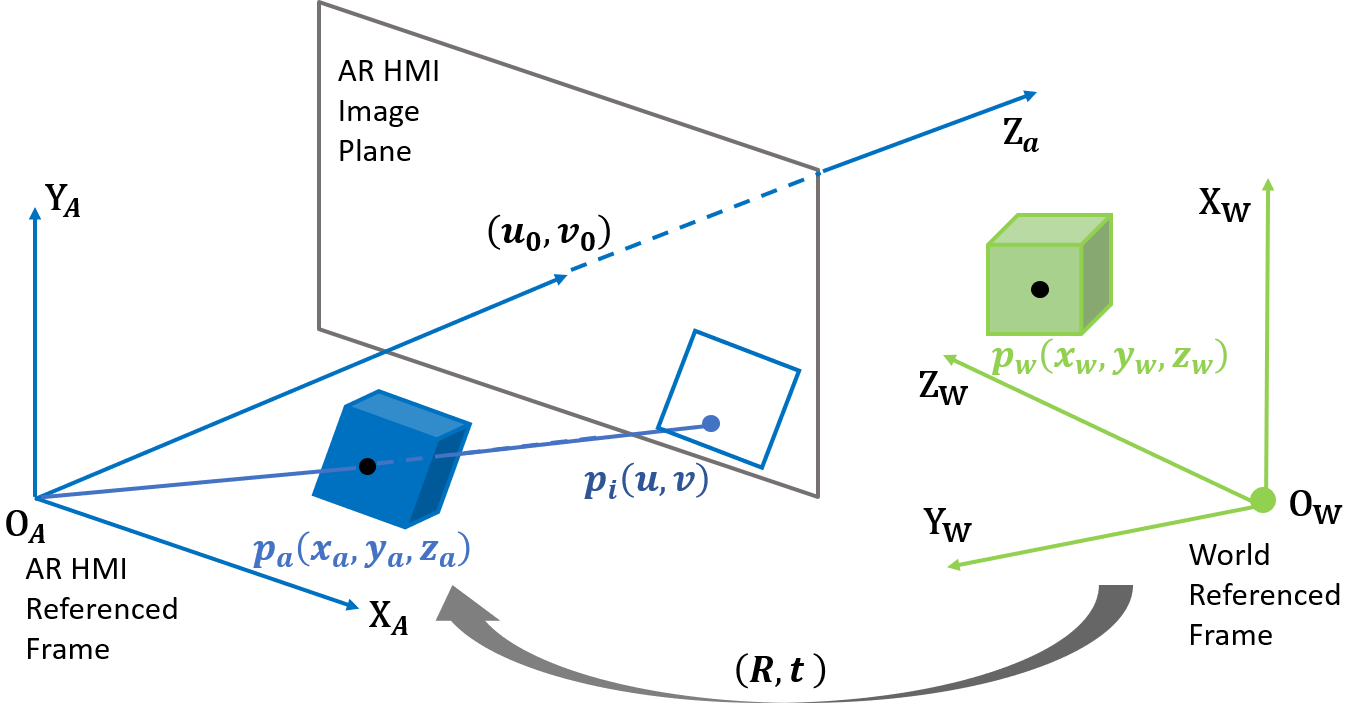}
    \caption{Coordinate transformation of the slot from the world referenced frame to AR HMI referenced frame}
    \label{transform}
\end{figure}

The AR HMI is displayed on an image plane (e.g., windshield) through the projector unit, where a front-view camera is needed to identify the road geometry, so a slot can be correctly overlaid on the road surface from the driver's field-of-view. In order to transform the slot from its global position and size (calculated in our control module) to the AR HMI, we develop a coordinate transformation algorithm based on the pinhole camera projection model, which is illustrated in Fig. \ref{transform}. 

The extrinsic parameter matrix in this algorithm identifies the transformation between the world referenced frame and the AR HMI reference frame. It consists of a $3\times3$ rotation matrix $R$ and a $3\times1$ translation vector $t$. Given a 3D point of the slot in the world referenced frame $p_w(x_w, y_w, z_w)$, its corresponding point $p_a$ in the AR referenced frame can be calculated as

\begin{equation}
   p_a = \begin{bmatrix} R & t \end{bmatrix}_{3 \times 4} \begin{pmatrix} x_w\\y_w\\z_w\\1 \end{pmatrix}_{4 \times 1}
\end{equation}

Then, the intrinsic parameter matrix is applied, which contains the parameters of the AR HMI's projection device, such as the focal length and lens distortion. Let ($u_0, v_0$) be the coordinates of the principle point of the image plane (i.e., image center), $d_x$ and $d_y$ be the physical size of pixels, and $f$ be the focal length, the projected point $p_i(u,v)$ on the AR HMI image plane can be calculated as

\begin{equation}
   p_i = 
   \begin{bmatrix} z_a d_x / f & 0 &  -z_a d_x u_0/ f \\
   0 & z_a d_y/ f & -z_a d_y v_0/ f\\
   0 & 0 & z_a \end{bmatrix}_{3 \times 3} \begin{pmatrix} p_a \end{pmatrix}_{3 \times 1}
\end{equation}

Therefore, given the position and size of any slot in the world referenced frame, we are able to transform the slot to the AR HMI image plane, so it can be properly shown to the driver of the vehicle. The calculation of slot's position and size will be discussed in the control module of the next subsection.

\subsection{Planning and Control of Connected Vehicles}
The planning module and the control module of this AR-based ADAS are developed to provide the inputs for the aforementioned AR HMI. As we design the AR HMI in an intuitive manner that visualizes vehicles as slots projected along the roadway, we first need our planning module to reserve those slots for different crossing vehicles, and then use our control module to adjust the appropriate sizes and speeds of the slots based on the vehicles' real-time information.

\subsubsection{Planning Module}
As illustrated in Fig. \ref{Arc}, the planning module takes as input the current status of the ego vehicle while approaching the intersection, and schedules its sequence of crossing the intersection by querying the slot pool. Once its slot is reserved, the ego vehicle can connect and cooperate with its reference vehicles that have conflicting paths with itself. The slot reservation algorithm (\textit{Algorithm 1}) is developed below for the planning module of the ADAS.

\begin{algorithm}
\small
  
\SetAlgoLined
\KwData{Ego vehicle $i$'s path $p_i$ from the current link to the next link, $i$'s longitudinal position $r_i$, $i$'s longitudinal speed $v_i$, $i$'s longitudinal acceleration $a_i$, car-following time headway $t_h$, reservation-trigger time constant $t_\theta$, reservation-trigger geo-fence distance constant $d_\theta$}
\KwResult{Ego vehicle $i$'s reserved slot $s_i$, $i$'s reference vehicles $j$s}
Ego vehicle $i$ enters the current link\;

\While{$i$ is not assigned $s_i$ at the current intersection}{
Calculate estimated time of arrival (ETA) $t_i = f(r_i, v_i, a_i)$\;

\If{There is/will be an immediate preceding vehicle $j$ on the same lane}{
Update ETA $t_i = min(t_i, t_j + t_h)$\;
}

Calculate distance to arrival $d_i$ based on $r_i$;

\If{$t_i <= t_\theta$ $||$ $d_i <= d_\theta$}{
Query conflicting vehicle $j$ whose path\newline $p_j \cap p_i$ $!=$ $\emptyset$ \;
Query the slot pool regarding the maximum slot number of all conflicting vehicles $s_j^{max}$\;
Assign slot number to the ego vehicle $s_i = s_j^{max} + 1$\;
Connect conflicting vehicle $j$ as a reference vehicle of $i$\;
}
}
\While{$i$ leaves the current link}{
Reset the slot number $s_i = 0$\;
Disconnect all reference vehicles\;}

\caption{\small{Slot reservation for crossing vehicles at an unsignalized intersection}}
\end{algorithm}

Instead of adopting a first-come-first-served policy \cite{dresner2008amultiagent}, which simply assigns a vehicle with a slot when it enters a pre-defined geo-fence, we develop a new slot reservation algorithm that accounts for various statuses of a vehicle. The estimated time of arrival (ETA) of a vehicle is considered, which quantifies a specific point in time when a vehicle is supposed to cross the intersection. Due to the limitation of space, the calculation of ETA is not covered in this subsection, which can be referred to our previous work \cite{wang2018distributed}. However, an additional step in our slot reservation algorithm is that, this ETA value of a vehicle is further updated by the ETA value of any immediate preceding vehicle of this vehicle. This means the traffic condition is also considered in the correct calculation of ETA, since a following vehicle's ETA cannot be earlier than its preceding vehicle's ETA, where a constant car-following time headway $t_h$ must be guaranteed as the delay in between.

Therefore, the condition that triggers a vehicle's slot reservation request is two-fold: Either its ETA is lower than a predefined time constant $t_\theta$, or it enters a pre-defined geo-fence of this intersection. This prevents some corner cases such as an ego vehicle enters the geo-fence earlier, but has a much lower speed (i.e., expected to arrive at the intersection later) than its conflicting vehicle coming from another direction. In that case, the conflicting vehicle needs to significantly decelerate to follow the ego vehicle to cross the intersection. Once \textit{Algorithm 1} is implemented, such unnecessary speed adjustments will not exsit anymore, and the overall traffic throughput and energy efficiency will be improved at this intersection.

\subsubsection{Control Module}
Once a connected vehicle is assigned a slot and connected with its reference vehicles, vehicle information is constantly transmitted among them. The control module of the ego vehicle aims to adjust the positions and sizes of its reference vehicles' reserved slots, so the slots can be better visualized for the AR HMI.

First, a target speed of the ego vehicle is calculated, based on the information received from its leading reference vehicle $j$, whose reserved slot is right in front of the ego vehicle's slot (i.e., $s_j = s_i - 1$). This target speed allows the ego vehicle to follow the movement of its reference vehicle's slot with the car-following time headway. A feedforward/feedback control algorithm is developed to calculate this target speed (that needs to be executed at the next time step) $v_i(t + \delta t)$. The feedback consensus control part is written as follow

\begin{multline} \label{v_sug}
v_i(t + \delta t) = v_i(t) + \Bigg[-\alpha_{ij} k_{ij} \cdot 
\bigg[\Big(r_i(t) - r_j\big(t - \tau_{ij}(t)\big) + v_i(t)\\
 \cdot \big(t_h + \tau_{ij}(t)\big)\Big) + \gamma_i \cdot \Big(v_i(t) - v_j\big(t - \tau_{ij}(t)\big)\Big)\bigg]\Bigg] \cdot \delta t
\end{multline}

\noindent where $\delta t$ is the length of each time step, $ v_i(t)$ is the current longitudinal speed of the vehicle, $ r_i(t)$ is the current longitudinal position of the vehicle, $\alpha_{ij}$ denotes the value of adjacency matrix. The time-variant communication delay between two vehicles is denoted as $\tau_{ij}(t)$, which is assumed as a normal distribution in this study, with a mean value of 40 ms and a standard deviation of 0.0259 based on our test results \cite{wang2020adigital}. The control gains $k_{ij}$ and $\gamma_i$ in this feedback control algorithm can be either defined as constants, or further tuned by a feedforward control algorithm to guarantee the safety, efficiency, and comfort of this slot-following process. A lookup-table approach is adopted to dynamically calculate these control gains, based on the initial speeds of two vehicles, as well as their initial headway. In short, it can be summarized as

\begin{equation}
    \{k_{ij}, \gamma_i\} = f\big(v_i(0), v_j(0), r_i(0) - r_j(0)\big)
\end{equation}

\noindent where the details can be referred to our previous work \cite{wang2019lookup}.

This target speed value $v_i(t + \delta t)$ is directly fed into the automated controller of CAVs to control their longitudinal speed. As for the AR HMI on human-driven connected vehicles, this target speed is the input to calculate the positions and sizes of the slots reserved by reference vehicles, where \textit{Algorithm 2} is developed below for the control module of the ADAS.

\begin{algorithm}
\small
\SetAlgoLined
\KwData{Ego vehicle $i$'s target speed at the next time step $v_i(t + \delta t)$, $i$'s reference vehicle $j$, $i$'s longitudinal position $r_i$, $i$'s lateral position $x_i$, $j$'s longitudinal position $r_j$, car-following time headway $t_h$, $i$'s path $p_i$ from the current link to the next link, $j$'s path $p_j$ from the current link to the next link, $j$'s longitudinal position $r_j$, $j$'s length $l_j$, $j$'s width $w_j$}
\KwResult{$j$'s reserved slot's longitudinal position $r_{s_j}$, lateral position $x_{s_j}$, length $l_{s_j}$, width $w_{s_j}$}

\For{Ego vehicle $i$'s all reference vehicles $j$s, $j = 1,2,...,n$}{
Calculate the conflicting point $O_{ij}$ of $i$'s path and $j$'s path\;
Calculate the distance difference $\delta_{ij}$ from $i$'s lane and $j$'s lane to the conflicting point\;
Calculate distances to arrival $d_i$ and $d_j$ based on $r_i$ and $r_j$\;

\While{$i$ does not cross the conflicting point $O_{ij}$}{
Calculate $j$'s reserved slot's longitudinal position $r_{s_j} = r_i - (d_i - d_j) + \delta_{ij}$\;
Set $j$'s reserved slot's lateral position $x_{s_j} = x_i$\;
Calculate $j$'s reserved slot's width $w_{s_j} = w_j$\;
Calculate $j$'s reserved slot's length $l_{s_j} = max(l_j, v_i(t + \delta t) * t_h)$\;
}

\While{$i$ crosses the conflicting point $O_{ij}$}{
Reset the slot information $r_{s_j}, w_{s_j}, l_{s_j}$\;
}
}

\caption{\small{Slot adjustment for AR HMI visualization}}
\end{algorithm}

\section{Game Engine Modeling and Evaluation} \label{sec:Sim}
\subsection{Game Engine for Modeling Connected Vehicles}
Game engines enable the design of video games for software developers, which typically consist of a rendering engine for 2-D or 3-D graphics, a physics engine for collision detection and response, and a scene graph for the management of multiple elements (e.g., models, sound, scripting, threading, etc.). Along with the rapid development of game engines in recent years, their functions have been broadened to a wider scope: data visualization, training, medical, and military use. Game engines also become popular options in the development of advanced vehicular technology \cite{ma2020new}, which have been used to study driver behaviors \cite{wang2020driver}, prototype connected vehicle systems \cite{wang2019cooperative, liu2020sensor}, and simulate autonomous driving \cite{dosovitskiy2017carla, rong2020lgsvl}.

\begin{figure}[ht!]
    \centering
    \includegraphics[width=1.0\columnwidth]{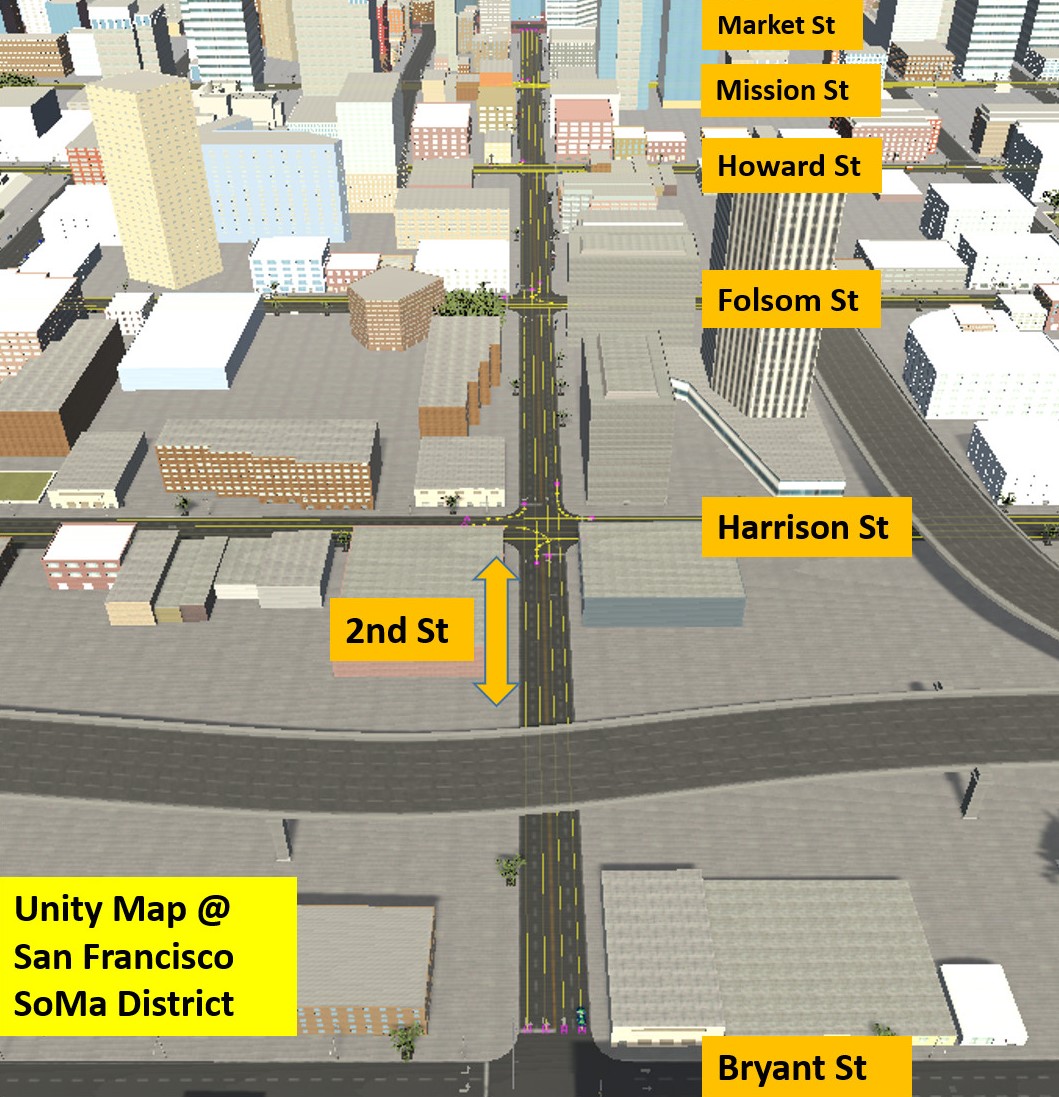}
    \caption{A four-intersection corridor HD map built in Unity game engine based on the SoMa district in San Francisco}
    \label{map}
\end{figure}

In this study, we adopt Unity game engine to conduct modeling and evaluation of our AR-based ADAS, given its advantages of graphics design and visualization, as well as its easiness to connect with external driving simulators \cite{unity}. As shown in Fig. \ref{map}, the map built by LGSVL is adopted in our study, which is based on the South of Market (SoMa) district in San Francisco \cite{rong2020lgsvl}. Shown as yellow lines on the road surface, centimeter-level routes along the 2nd Street, Harrison Street, Folsom Street, Howard Street, and Mission Street are further modeled in this study, so map matching and path planning features can be enabled in our ADAS. The planning and control modules we develop in this study are modeled on vehicles in this environment through Unity's C\# API. The AR HMI is also designed on the ego vehicle through Unity's visualization feature. 

\subsection{Human-in-the-Loop Simulation}
\begin{figure}[ht!]
    \centering
    \includegraphics[width=1.0\columnwidth]{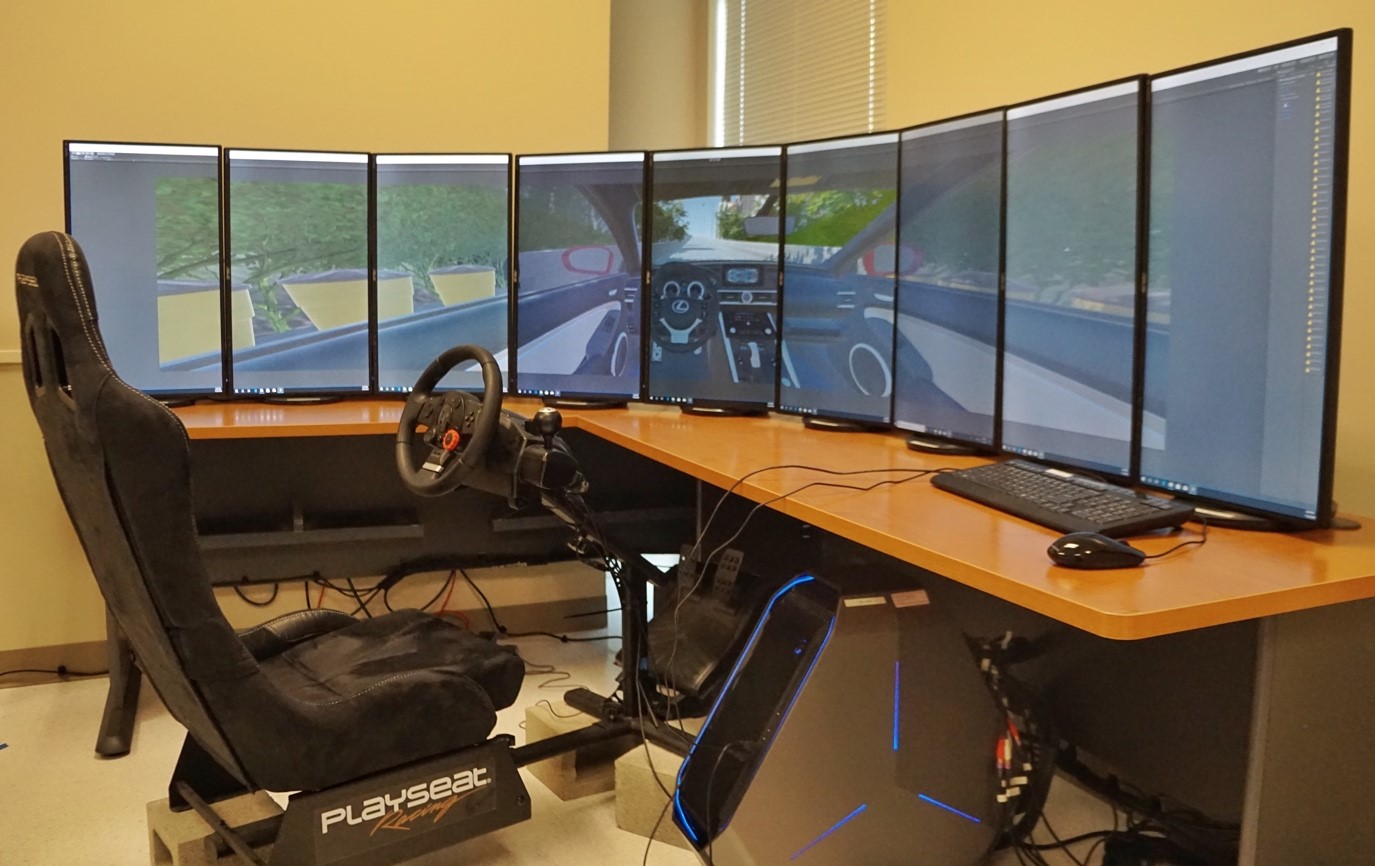}
    \caption{Driving simulator platform to conduct human-in-the-loop simulation}
    \label{Sim}
\end{figure}

To evaluate our proposed AR-based ADAS, we conduct HITL simulation with drivers controlling the external driving simulator. As shown in Fig. \ref{Sim} The driving simulator platform is built with a desktop (processor Intel Core i7-9750 @2.60GHz, memory 32.0 GB), a Logitech G29 Driving Force racing wheel, and Unity 2019.2.11f1.

The invited participants in this simulation are advised to drive the ego vehicle in the Unity environment, which travels along the 2nd Street. The ego vehicle starts from the Bryant Street with zero speed, and then crosses four consecutive unsignalized intersections from south to north. All other vehicles in the simulation are non-player characters (NPCs), which run the proposed planning and control modules as CAVs.

Additionally, all participants drive the ego vehicle in the baseline scenario, where all four intersections have traditional fixed-timing traffic signals. NPC vehicles are randomly generated from all directions at each intersection, and they are not enabled with any CAV feature in this scenario. This enables us to investigate the benefits brought by the proposed AR-based ADAS to the existing transportation systems.

\subsection{Simulation Results and Evaluation}
\begin{figure}[ht!]
    \centering
    \subfloat[]{%
    \includegraphics[width=1.0\columnwidth]{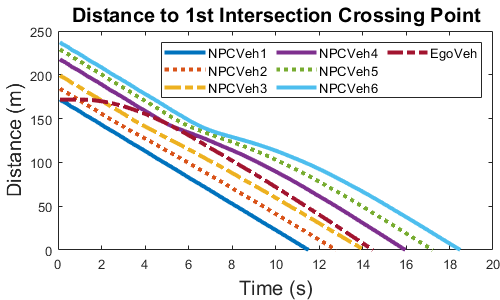}
    }\\
    \subfloat[]{%
    \includegraphics[width=1.0\columnwidth]{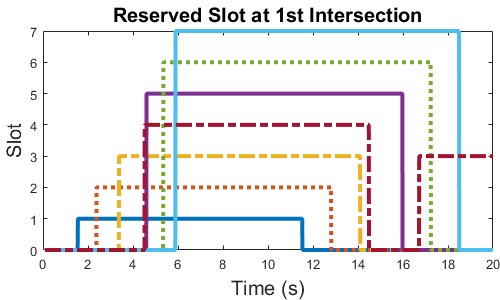}
    }\\
    \caption{A sample simulation result of applying AR-based ADAS at the 2nd St \& Harrison St intersection, where the ego vehicle is driven by a human driver on the driving simulation platform}
    \label{Result}
\end{figure}

Among all the trips conducted by the participants, one sample simulation result is shown in Fig. \ref{Result}. This sample result was generated when the participant drove the ego vehicle through unsignalized intersections, under the guidance of our proposed AR-based ADAS. Specifically, the first segment of the whole trip is picked out, where the ego vehicle crossed the first intersection (i.e., 2nd Street \& Harrison Street intersection) in collaboration with all other six NPC vehicles.

The distance-time plot in Fig. \ref{Result}(a) shows the ego vehicle was able to keep a relatively safe distance regarding its reference vehicles, including its immediate preceding NPC vehicle 3. Meanwhile, the ego vehicle also acts as a reference vehicle for NPC vehicle 4, 5 and 6, where NPC vehicle 4 considers the ego vehicle as its immediate preceding vehicle. Since the proposed feedforward/feedback control algorithm was applied to these three NPC vehicles, they consecutively decelerated during 5-12 s to maintain a relatively safe distance with their immediate preceding vehicles.

The process of all seven vehicles reserving slots is shown in Fig. \ref{Result}(b), which corresponds to the vehicle trajectories in Fig. \ref{Result}(a). Once a vehicle was assigned a slot by our proposed \textit{Algorithm 1}, they immediately identified their reference vehicles and apply the control algorithm. Once they crossed the current intersection, the reserved slots were reset to zero, waiting for new assignments while approaching the next intersection. It can be noticed from this plot that, the ego vehicle (i.e., dark-red dashed line) was assigned a new slot of three after it entered the next link, even before NPC vehicle 5 and 6 crossed the first intersection. This means the proposed slot reservation process is independent at each intersection, which continues running when different vehicles approaching and leaving the intersection.

\begin{figure}[ht!]
    \centering
    \includegraphics[width=1.0\columnwidth]{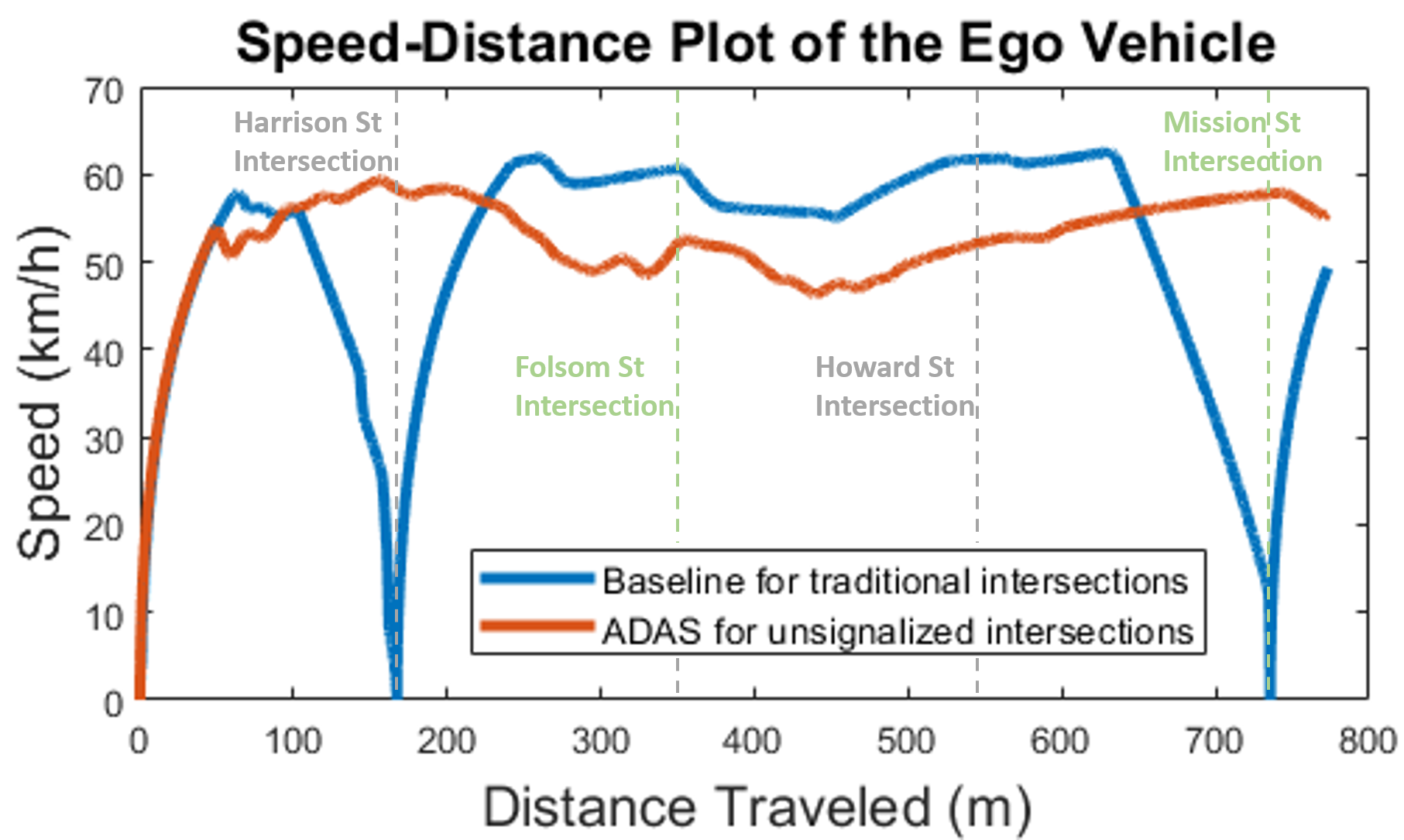}
    \caption{A sample comparison when the ego vehicle travels through the whole corridor 1) with traditional intersections and 2) with unsignalized intersections using the proposed ADAS}
    \label{DV}
\end{figure}

A sample comparison between the unsignalized intersection scenario and the baseline traditional intersection scenario is shown in Fig. \ref{DV}. This speed-distance plot was created when the same participant drove the ego vehicle through all four intersections in Fig. \ref{map}. In the baseline scenario, the ego vehicle ran into red lights at the first and the fourth intersections, while directly passed the second and the third intersections during green lights. In the unsignalized intersection scenario, however, the ego vehicle maintained a relatively stable speed while travelling through all four intersections, without any full stop at any intersection. Although a higher maximum speed was reached in the baseline scenario, the excessive speed changes significantly increased the travel time and the energy consumption. Based on all trips conducted by the participants in HITL simulation, an average of 20\% reduction in travel time, and an average of 23.7\% reduction in fuel consumption (calculated by the open-source MOVESTAR model \cite{wang2020movestar} and assuming all gasoline vehicles) can be achieved by applying the proposed AR-based ADAS.


\section{Conclusion and Future Work} \label{sec:Con}
In this study, an AR-based ADAS was designed for connected vehicles, which visualizes the guidance information to vehicle drivers in a more intuitive manner. Instead of making the HMI a simplified output of whatever is provided by other modules of the ADAS, we designed the planning and control modules of our ADAS to better serve the HMI, making this system human-centered. A slot reservation methodology was proposed in the unsignalized intersection use case, where a driver can cooperate with other crossing vehicles at intersections by simply following the guidance on the AR HMI. Modeling and evaluation of this AR-based ADAS were conducted in Unity game engine, where HITL simulation results proved the its benefits in travel time and energy consumption.

To take this study one step further, vulnerable road users such as pedestrians and bicycles need to be considered in the modeling and visualization process. The term ``mixed traffic environment'' does not only refer to an environment mixed with different kinds of vehicles, but also includes vulnerable road users that have the highest priority in the environment. How to build the ADAS of connected vehicles that could cooperate with vehicles, bicycles and pedestrians at the same time remains an interesting question to be solved.

\section*{Acknowledgment}
The contents of this paper only reflect the views of the authors, who are responsible for the facts and the accuracy of the data presented herein. The contents do not necessarily reflect the official views of Toyota Motor North America.

\ifCLASSOPTIONcaptionsoff
  \newpage
\fi

\bibliographystyle{IEEEtran}
\bibliography{Reference.bib}

\begin{thebibliography}{10}
\providecommand{\url}[1]{#1}
\csname url@samestyle\endcsname
\providecommand{\newblock}{\relax}
\providecommand{\bibinfo}[2]{#2}
\providecommand{\BIBentrySTDinterwordspacing}{\spaceskip=0pt\relax}
\providecommand{\BIBentryALTinterwordstretchfactor}{4}
\providecommand{\BIBentryALTinterwordspacing}{\spaceskip=\fontdimen2\font plus
\BIBentryALTinterwordstretchfactor\fontdimen3\font minus
  \fontdimen4\font\relax}
\providecommand{\BIBforeignlanguage}[2]{{%
\expandafter\ifx\csname l@#1\endcsname\relax
\typeout{** WARNING: IEEEtran.bst: No hyphenation pattern has been}%
\typeout{** loaded for the language `#1'. Using the pattern for}%
\typeout{** the default language instead.}%
\else
\language=\csname l@#1\endcsname
\fi
#2}}
\providecommand{\BIBdecl}{\relax}
\BIBdecl

\bibitem{wang2020asurvey}
Z.~{Wang}, Y.~{Bian}, S.~E. {Shladover}, G.~{Wu}, S.~E. {Li}, and M.~J.
  {Barth}, ``A survey on cooperative longitudinal motion control of multiple
  connected and automated vehicles,'' \emph{IEEE Intelligent Transportation
  Systems Magazine}, vol.~12, no.~1, pp. 4--24, 2020.

\bibitem{monreal2016human}
C.~{Olaverri-Monreal} and T.~{Jizba}, ``Human factors in the design of
  human–machine interaction: An overview emphasizing v2x communication,''
  \emph{IEEE Transactions on Intelligent Vehicles}, vol.~1, no.~4, pp.
  302--313, 2016.

\bibitem{wang2019early}
Z.~{Wang}, Y.~{Hsu}, A.~{Vu}, F.~{Caballero}, P.~{Hao}, G.~{Wu},
  K.~{Boriboonsomsin}, M.~J. {Barth}, A.~{Kailas}, P.~{Amar}, E.~{Garmon}, and
  S.~{Tanugula}, ``Early findings from field trials of heavy-duty truck
  connected eco-driving system,'' in \emph{2019 IEEE Intelligent Transportation
  Systems Conference (ITSC)}, 2019, pp. 3037--3042.

\bibitem{campbell2016human}
J.~Campbell, J.~Brown, J.~Graving, C.~Richard, M.~Lichty, T.~Sanquist, and
  J.~Morgan, ``Human factors design guidance for driver-vehicle interfaces,''
  {U.S. Department of Transportation National Highway Traffic Safety
  Administration}, Tech. Rep., 2016.

\bibitem{kenney2011dedicated}
J.~B. Kenney, ``Dedicated short-range communications (dsrc) standards in the
  united states,'' \emph{Proceedings of the IEEE}, vol.~99, no.~7, pp.
  1162--1182, July 2011.

\bibitem{chen2017vehicle}
S.~Chen, J.~Hu, Y.~Shi, Y.~Peng, J.~Fang, R.~Zhao, and L.~Zhao,
  ``Vehicle-to-everything ({V2X}) services supported by {LTE}-based systems and
  {5G},'' \emph{IEEE Communications Standards Magazine}, vol.~1, no.~2, pp.
  70--76, 2017.

\bibitem{altan2017glidepath}
O.~D. Altan, G.~Wu, M.~J. Barth, K.~Boriboonsomsin, and J.~A. Stark,
  ``Glidepath: Eco-friendly automated approach and departure at signalized
  intersections,'' \emph{IEEE Transactions on Intelligent Vehicles}, vol.~2,
  no.~4, pp. 266--277, Dec 2017.

\bibitem{conceicao2013virtual}
H.~{Conceição}, M.~{Ferreira}, and P.~{Steenkiste}, ``Virtual traffic lights
  in partial deployment scenarios,'' in \emph{2013 IEEE Intelligent Vehicles
  Symposium (IV)}, 2013, pp. 988--993.

\bibitem{monreal2010theseethrough}
C.~{Olaverri-Monreal}, P.~{Gomes}, R.~{Fernandes}, F.~{Vieira}, and
  M.~{Ferreira}, ``The see-through system: A vanet-enabled assistant for
  overtaking maneuvers,'' in \emph{2010 IEEE Intelligent Vehicles Symposium},
  2010, pp. 123--128.

\bibitem{bertini2006dynamics}
\BIBentryALTinterwordspacing
R.~L. Bertini, S.~Boice, and K.~Bogenberger, ``Dynamics of variable speed limit
  system surrounding bottleneck on german autobahn,'' \emph{Transportation
  Research Record}, vol. 1978, no.~1, pp. 149--159, 2006. [Online]. Available:
  \url{https://doi.org/10.1177/0361198106197800119}
\BIBentrySTDinterwordspacing

\bibitem{nygardhs2007variable}
S.~Nyg{\aa}rdhs and G.~Helmers, ``{VMS} - {Variable Message Signs}: A
  literature review,'' Infrastructure maintenance, Tech. Rep. 570A, 2007.

\bibitem{quinlan2010bringing}
M.~Quinlan, T.-C. Au, J.~Zhu, N.~Stiurca, and P.~Stone, ``Bringing simulation
  to life: A mixed reality autonomous intersection,'' in \emph{Proceedings of
  IEEE/RSJ International Conference on Intelligent Robots and Systems (IROS)},
  October 2010.

\bibitem{feng2018anaugmented}
Y.~{Feng}, C.~{Yu}, S.~{Xu}, H.~X. {Liu}, and H.~{Peng}, ``An augmented reality
  environment for connected and automated vehicle testing and evaluation*,'' in
  \emph{2018 IEEE Intelligent Vehicles Symposium (IV)}, 2018, pp. 1549--1554.

\bibitem{wang2020driver}
Z.~{Wang}, X.~{Liao}, C.~{Wang}, D.~{Oswald}, G.~{Wu}, K.~{Boriboonsomsin},
  M.~{Barth}, K.~{Han}, B.~{Kim}, and P.~{Tiwari}, ``Driver behavior modeling
  using game engine and real vehicle: A learning-based approach,'' \emph{IEEE
  Transactions on Intelligent Vehicles}, pp. 1--1, 2020.

\bibitem{continentalAR}
\BIBentryALTinterwordspacing
Continental, ``Augmented-reality {HUD},'' Accessed: 2020-07-06. [Online].
  Available:
  \url{https://www.continental-automotive.com/en-gl/Passenger-Cars/Information-Management/Head-Up-Displays/Augmented-Reality-HUD}
\BIBentrySTDinterwordspacing

\bibitem{vwAR}
\BIBentryALTinterwordspacing
Volkswagen, ``{AR Head-Up} display,'' Accessed: 2020-07-06. [Online].
  Available:
  \url{https://www.volkswagen.co.uk/electric/id/id-family/id3.html#comfort}
\BIBentrySTDinterwordspacing

\bibitem{fhwa2014signalized}
\BIBentryALTinterwordspacing
{U.S. Department of Transportation Federal Highway Administration},
  ``Signalized intersections: An informational guide,'' 2020-06-02. [Online].
  Available:
  \url{https://safety.fhwa.dot.gov/intersection/conventional/signalized/fhwasa13027/ch1.cfm}
\BIBentrySTDinterwordspacing

\bibitem{dresner2008amultiagent}
K.~Dresner and P.~Stone, ``A multiagent approach to autonomous intersection
  management,'' \emph{Journal of artificial intelligence research}, vol.~31,
  pp. 591--656, 2008.

\bibitem{neuendorf2004thevehicle}
N.~Neuendorf and T.~Bruns, ``The vehicle platoon controller in the
  decentralised, autonomous intersection management of vehicles,'' in
  \emph{Proceedings of the IEEE International Conference on Mechatronics, 2004.
  ICM '04.}, Jun. 2004, pp. 375--380.

\bibitem{jin2013platoon}
Q.~Jin, G.~Wu, K.~Boriboonsomsin, and M.~Barth, ``Platoon-based multi-agent
  intersection management for connected vehicle,'' in \emph{2013 16th
  International IEEE Conference on Intelligent Transportation Systems (ITSC))},
  Oct. 2013, pp. 1462--1467.

\bibitem{xu2018distributed}
B.~Xu, S.~E. Li, Y.~Bian, S.~Li, X.~J. Ban, J.~Wang, and K.~Li, ``Distributed
  conflict-free cooperation for multiple connected vehicles at unsignalized
  intersections,'' \emph{Transportation Research Part C: Emerging
  Technologies}, vol.~93, pp. 322--334, 2018.

\bibitem{wang2018distributed}
\BIBentryALTinterwordspacing
Z.~Wang, G.~Wu, and M.~Barth, ``Distributed consensus-based cooperative highway
  on-ramp merging using {V2X} communications,'' in \emph{SAE Technical Paper},
  Apr. 2018. [Online]. Available: \url{https://doi.org/10.4271/2018-01-1177}
\BIBentrySTDinterwordspacing

\bibitem{wang2020adigital}
Z.~Wang, X.~Liao, X.~Zhao, K.~Han, P.~Tiwari, M.~J. Barth, and G.~Wu, ``A
  digital twin paradigm: {Vehicle-to-Cloud} based advanced driver assistance
  systems,'' in \emph{2020 IEEE 91st Vehicular Technology Conference}, May
  2020, pp. 1--6.

\bibitem{wang2019lookup}
Z.~Wang, K.~Han, B.~Kim, G.~Wu, and M.~J. Barth, ``Lookup table-based consensus
  algorithm for real-time longitudinal motion control of connected and
  automated vehicles,'' \emph{arXiv:1902.07747v2}, 2019.

\bibitem{ma2020new}
J.~Ma, C.~Schwarz, Z.~Wang, M.~Elli, G.~Ros, and Y.~Feng, ``New simulation
  tools for training and testing automated vehicles,'' in \emph{Road Vehicle
  Automation 7}, G.~Meyer and S.~Beiker, Eds.\hskip 1em plus 0.5em minus
  0.4em\relax Cham: Springer International Publishing, 2020, pp. 111--119.

\bibitem{wang2019cooperative}
Z.~Wang, G.~Wu, K.~Boriboonsomsin, M.~Barth \emph{et~al.}, ``Cooperative ramp
  merging system: Agent-based modeling and simulation using game engine,''
  \emph{SAE International Journal of Connected and Automated Vehicles}, vol.~2,
  no.~2, 2019.

\bibitem{liu2020sensor}
Y.~Liu, Z.~Wang, K.~Han, Z.~Shou, P.~Tiwari, and J.~H.~L. Hansen, ``Sensor
  fusion of camera and cloud digital twin information for intelligent
  vehicles,'' in \emph{IEEE Intelligent Vehicles Symposium (IV)}, Jun. 2020.

\bibitem{dosovitskiy2017carla}
A.~Dosovitskiy, G.~Ros, F.~Codevilla, A.~Lopez, and V.~Koltun, ``{CARLA}: An
  open urban driving simulator,'' \emph{arXiv preprint arXiv:1711.03938}, 2017.

\bibitem{rong2020lgsvl}
G.~Rong, B.~H. Shin, H.~Tabatabaee, Q.~Lu, S.~Lemke, M.~Mo{\v{z}}eiko,
  E.~Boise, G.~Uhm, M.~Gerow, S.~Mehta \emph{et~al.}, ``Lgsvl simulator: A high
  fidelity simulator for autonomous driving,'' \emph{arXiv preprint
  arXiv:2005.03778}, 2020.

\bibitem{unity}
\BIBentryALTinterwordspacing
{Unity}, ``Unity for all.'' [Online]. Available: \url{https://unity.com/}
\BIBentrySTDinterwordspacing

\bibitem{wang2020movestar}
Z.~Wang, G.~Wu, and G.~Scora, ``{MOVESTAR}: An open-source vehicle fuel and
  emission model based on usepa moves,'' 2020.

\end{thebibliography}

\vskip 0pt plus -1fil
\begin{IEEEbiography}
[{\includegraphics[width=1in,height=1.25in,clip,keepaspectratio]{authors//Ziran.jpg}}]
{Ziran Wang}
(S'16-M'19) received the Ph.D. degree in mechanical engineering from University of California at Riverside in 2019, and the B.E. degree in mechanical engineering and automation from Beijing University of Posts and Telecommunications in 2015, respectively. He is currently a Research Scientist at Toyota Motor North America, InfoTech Labs. His research focuses on connected and automated vehicle technology, including motion control and vehicle-to-everything (V2X) communication.
\end{IEEEbiography}

\vskip 0pt plus -1fil
\begin{IEEEbiography}
[{\includegraphics[width=1in,height=1.25in,clip,keepaspectratio]{authors//KT.jpg}}]
{Kyungtae (KT) Han}
(M'97-SM'15) received the Ph.D. degree in electrical and computer engineering from The University of Texas at Austin in 2006. He is currently a Principal Researcher at Toyota Motor North America, InfoTech Labs. Prior to joining Toyota, Dr. Han was a Research Scientist at Intel Labs, and a Director in Locix Inc. His research interests include cyber-physical systems, connected and automated vehicle technique, and intelligent transportation systems.  
\end{IEEEbiography}

\vskip 0pt plus -1fil
\begin{IEEEbiography}
[{\includegraphics[width=1in,height=1.25in,clip,keepaspectratio]{authors//Prashant.jpg}}]
 {Prashant Tiwari}
received the Ph.D. degree in mechanical engineering from Rensselaer Polytechnic Institute in 2004, and the MBA degree from University of Chicago in 2016. He is currently a Executive Director at Toyota Motor North America, InfoTech Labs. Dr. Tiwari is highly active in Automotive Edge Computing Consortium (AECC) and SAE. Prior to joining Toyota, Dr. Tiwari held several leadership positions of increasing responsibilities at GE and UTC Aerospace Systems.  
\end{IEEEbiography}

\end{document}